\documentclass[12pt,letterpaper]{JHEP3}

\usepackage{amsmath,amsfonts,amscd,graphicx}

\title{Series Solution and Minimal Surfaces in AdS}

\author{Antal Jevicki, Kewang Jin \\ Department of Physics, Brown University, \\ Box 1843, Providence, RI 02912, USA \\ E-mail: \email{antal@het.brown.edu,jin@het.brown.edu}}

\abstract{According to the Alday-Maldacena program the strong coupling limit of Super Yang-Mills scattering amplitudes is given by minimal area surfaces in AdS spacetime with a boundary consisting of a momentum space polygon. The string equations in AdS systematically reduce to coupled Toda type equations whose Euclidean classical solutions are then of direct relevance. While in the simplest case of AdS$_3$ exact solutions were known from earlier studies of the sinh-Gordon equation, there exist at present no similar exact forms for the generalized Toda equations related to AdS$_d$ with $d \ge 4$. In this paper we develop a series method for the solution to those equations and evaluate their contribution to the finite piece of the worldsheet area. For the known sinh-Gordon case the method is seen to give results in excellent agreement with the exact answer.}


\preprint{\tt{BROWN-HET-1588}}

\begin{document}

\section{Introduction}

The Gauge/String correspondence is characterized by the fact that the semiclassical limit \cite{Gubser:2002tv} of the AdS string is capable of providing answers for the strong coupling regime of gauge theory. This has been most successfully demonstrated for the case of ${\cal N}=4$ Super Yang-Mills theory where an exact Bethe ansatz solution is available \cite{Beisert:2006ez}, bridging the weak and strong couplings. Configurations of spinning \cite{Tseytlin:2003ii} and spiky \cite{Kruczenski:2004wg} strings were seen to match up perfectly with higher spin states of Yang-Mills theory.

The most explicit implementation of the correspondence was given by Alday and Maldacena \cite{Alday:2007hr}, through the Wilson loop representation \cite{Alday:2008yw} of Yang-Mills theory scattering amplitudes. Amplitudes in Yang-Mills theory have been studied extensively in weak coupling perturbation theory with the conjectured BDS proposal \cite{Bern:2005iz} (see also \cite{Anastasiou:2003kj}) in which the cusp anomaly contribution plays a central role. The representation of Alday and Maldacena \cite{Alday:2007hr} realizes the scattering amplitudes through Wilson loops with lightlike closed polygon boundaries, whose segments are given by the gluon momenta (see \cite{Mironov:2007qq,Dobashi:2008ia} for developments). Through this representation the strong coupling answer is obtained by evaluating minimal surfaces in AdS.

One way to approach the classical problem of constructing string solutions in AdS spacetime (in conformal gauge) is through a Pohlmeyer reduction \cite{Pohlmeyer,Jevicki:2007aa} of the associated classical nonlinear sigma model. This technique was applied previously to the construction of solutions in de Sitter spacetime \cite{de Vega} and extended to soliton and spiky Minkowski worldsheet solutions in a series of papers \cite{Jevicki:2008mm}. For the case of minimal surfaces with Euclidean worldsheet extension is applicable. The Pohlmeyer reduction reduces the nonlinear sigma model equations to a coupled system consisting of integrable Toda type equations and a conformal pair obeying the Cauchy-Riemann conditions. In Minkowski case the integrability of the (Toda type) theories provides (singular) soliton type solutions which were identified with spikes in \cite{Jevicki:2007aa,Jevicki:2008mm} (see also \cite{Berkovits:2008ic,Dorn:2009kq}). For the case of minimal surfaces in AdS$_3$, the Euclidean system can be seen to be identical to the SU(2) Hitchin equations on the associated Riemann surface. In a remarkable paper \cite{Alday:2009ga}, Alday and Maldacena have demonstrated that the highly nontrivial problem of matching the polygon boundary conditions is the same one as the mathematical problem of wall-crossing in the Hitchin system, which was recently studied in \cite{Gaiotto:2008cd} and earlier mathematical literature. This allowed for a beautiful evaluation of the eight point scattering amplitude in AdS$_3$. 

In the evaluation the curvature contribution to the minimal surface is given by an instanton-type solution of the Euclidean
sinh-Gordon equation. The construction of the sinh-Gordon solution is itself nontrival, its asymptotic expansion was accomplished
in an earlier work of McCoy, Tracy and Wu \cite{TTWu} through a mapping to a Painlev\'{e} equation. For this case, an exact multi-integral formula is also known, which allows for a weak and strong coupling expansion given in the work of Zamolodchikov \cite{Zamolodchikov:1994uw} (see also \cite{CFYG}). There exists as yet, no similar representation for the case of generalized Toda equations that appear in the case of AdS$_d$ for $d=4,5$ that would be of interest.

In this paper we intend to fill this gap by presenting a series method that will be applicable to equations associated with AdS$_d$, any dimensions. First we describe the method in the sinh-Gordon example, where comparing with the exact result one can judge
the accuracy of the technique. It is seen that with a few terms in the series one achieves a 99.85\% agreement (six points) for the area of the surface. We then describe the calculation for the $B_2$ Toda system associated with AdS$_4$. The method is based on an asymptotic large distance expansion, with a nontrivial matching at short distance giving nonlinear constraint equations on the coefficients in the expansion. The technique was used in the case of solitons in \cite{Manton:1978gf}.

The content of the paper goes as follows. In section two, we review the Pohlmeyer reduction of classical strings in AdS with Euclidean worldsheet. In section three, we give the series solution to the sinh-Gordon equation, evaluate the area and compare with the exact results. Then we generalize the method to the Toda equations in section four. While in the appendices, we summarize the exact solutions of the sinh-Gordon and kink cases.

\section{Pohlmeyer reduction of classical strings in AdS}

In general, sting equations in AdS$_d$ spacetime (in conformal gauge) are described by the non-compact nonlinear sigma model on $SO(d-1,2)$. Defining the AdS$_d$ space as $Y^2=-Y_{-1}^2-Y_0^2+Y_1^2+\cdots+Y_{d-1}^2=-1$, the Euclidean action reads
\begin{equation}
S_E={\sqrt{\lambda} \over 4\pi}\int d\tau d\sigma \Bigl(\partial_\mu Y \partial^\mu Y + \lambda (\sigma ,\tau)(Y \cdot Y+1)\Bigr),
\end{equation}
where $\tau,\sigma$ are the Euclidean worldsheet coordinates, the equations of motion are
\begin{equation}
\partial \bar{\partial} Y-(\partial Y \cdot \bar{\partial} Y) Y=0,
\end{equation}
with $z=(\sigma+i\tau)/2,\bar{z}=(\sigma-i\tau)/2$ and $\partial=\partial_\sigma-i\partial_\tau,\bar{\partial}=\partial_\sigma+i\partial_\tau$. In addition to guarantee the conformal gauge we have to impose the Virasoro conditions
\begin{equation}
\partial Y \cdot \partial Y = \bar{\partial} Y \cdot \bar{\partial} Y=0.
\end{equation}

It was demonstrated a number of years ago (by Pohlmeyer \cite{Pohlmeyer}) that nonlinear sigma models subject to Virasoro type constraints can be reduced to integrable field equations of sinh-Gordon (or Toda) type. This reduction is accomplished by concentrating on $SO(d-1,2)$ invariant sub-dynamics of the sigma model. The steps of the reduction were well described in \cite{de Vega,Barbashov:1982qz} and consist in the following. One starts by identifying first an appropriate set of basis vectors for the string coordinates
\begin{equation}
e_i=(Y,\partial Y,\bar{\partial} Y,B_4,\cdots,B_{d+1}), \qquad i=1,2,\cdots,d+1,
\end{equation}
where $B_i$ form an orthonormal set $B_i \cdot B_j=\delta_{ij},B_i \cdot Y=B_i \cdot \partial Y=B_i \cdot \bar{\partial} Y=0$. Defining the scalar field
\begin{equation}
\alpha(z,\bar{z}) \equiv \ln[\partial Y \cdot \bar{\partial} Y],
\end{equation}
one can derive the equation of motion for $\alpha$ has the form
\begin{equation}
\partial \bar{\partial} \alpha-e^{\alpha}-e^{-\alpha}\sum_{i=4}^{d+1} u_i v_i=0,
\label{eomalpha}
\end{equation}
where we have defined two sets of auxiliary fields
\begin{equation}
u_i \equiv B_i \cdot \partial^2 Y, \qquad v_i \equiv B_i \cdot \bar{\partial}^2 Y.
\end{equation}

This is an integrable system, the Lax pair can be found by expressing the derivatives of the basis vectors in terms of the basis itself
\begin{equation}
\partial e_i=A_{ij} e_j, \qquad \bar{\partial} e_i=B_{ij} e_j,
\end{equation}
where the matrices $A$ and $B$ are found to be
\begin{equation}
A=\begin{pmatrix} 0 & 1 & 0 & 0 & \cdots & 0 \cr 0 & \partial\alpha & 0 & u_4 & \cdots & u_{d+1} \cr e^{\alpha} & 0 & 0 & 0 & \cdots & 0 \cr 0 & 0 & -u_4 e^{-\alpha} & 0 & \cdots & B_{d+1} \cdot \partial B_4 \cr \vdots & \vdots & \vdots & \vdots & B_j \cdot \partial B_i & \vdots \cr 0 & 0 & -u_{d+1}e^{-\alpha} & B_4 \cdot \partial B_{d+1} & \cdots & 0 \end{pmatrix},
\end{equation}
\begin{equation}
B=\begin{pmatrix} 0 & 0 & 1 & 0 & \cdots & 0 \cr e^\alpha & 0 & 0 & 0 & \cdots & 0 \cr 0 & 0 & \bar{\partial}\alpha & v_4 & \cdots & v_{d+1} \cr 0 & -v_4 e^{-\alpha} & 0 & 0 & \cdots & B_{d+1} \cdot \bar{\partial}B_4 \cr \vdots & \vdots & \vdots & \vdots & B_j \cdot \bar{\partial}B_i & \vdots \cr 0 & -v_{d+1}e^{-\alpha} & 0 & B_4 \cdot \bar{\partial} B_{d+1} & \cdots & 0 \end{pmatrix}.
\end{equation}
The zero curvature condition
\begin{equation}
\bar{\partial} A-\partial B+[A,B]=0,
\label{eqn427}
\end{equation}
implies the equations of motion for $\alpha$ (\ref{eomalpha}) and for the auxiliary fields $u_i,v_i$
\begin{eqnarray}
\bar{\partial}u_i&=&\sum_{j \neq i}(B_j \cdot \bar{\partial}B_i) u_j, \\
\partial v_i&=&\sum_{j \neq i}(B_j \cdot \partial B_i) v_j.
\end{eqnarray}
Next, we are going to discuss two particular examples.

\subsection{The AdS$_3$ case}

In this case, the $A,B$ matrices are
\begin{equation}
A=\begin{pmatrix} 0 & 1 & 0 & 0 \cr 0 & \partial \alpha & 0 & u \cr e^\alpha & 0 & 0 & 0 \cr 0 & 0 & -u e^{-\alpha} & 0 \end{pmatrix},\qquad B=\begin{pmatrix} 0 & 0 & 1 & 0 \cr e^\alpha & 0 & 0 & 0 \cr 0 & 0 & \bar{\partial}\alpha & v \cr 0 & -v e^{-\alpha} & 0 & 0 \end{pmatrix}.
\end{equation}
The zero curvature condition (\ref{eqn427}) implies
\begin{equation}
\partial \bar{\partial} \alpha-e^\alpha-u(z) v(\bar{z}) e^{-\alpha}=0.
\label{eqn41}
\end{equation}
This is called the generalized sinh-Gordon equation. The fourth vector $B_4$ here is purely imaginary so that we can choose
\begin{equation}
u(z)=p(z), \qquad v(\bar{z})=-\bar{p}(\bar{z}).
\end{equation}
Using the reparametrization invariance on the worldsheet, one can make a shift to the scalar field
\begin{equation}
\alpha(z,\bar{z})=\hat{\alpha}(z,\bar{z})+{1 \over 2}\ln[p(z)\bar{p}(\bar{z})],
\label{change1}
\end{equation}
along with a change of variables as
\begin{equation}
w=\int \sqrt{p(z)}dz, \qquad \bar{w}=\int \sqrt{\bar{p}(\bar{z})}d\bar{z},
\label{change2}
\end{equation}
such that (\ref{eqn41}) reduces to the standard sinh-Gordon equation
\begin{equation}
\partial_w \bar{\partial}_{\bar{w}} \hat{\alpha}(w,\bar{w})-2 \sinh \hat{\alpha}=0.
\label{sinh}
\end{equation}

\subsection{The AdS$_4$ case}

For $d=4$, the $A,B$ matrices are
\begin{equation}
A=\begin{pmatrix} 0 & 1 & 0 & 0 & 0 \cr 0 & \partial\alpha & 0 & u_4 & u_5 \cr e^\alpha & 0 & 0 & 0 & 0 \cr 0 & 0 & -u_4 e^{-\alpha} & 0 & -\partial\gamma \cr 0 & 0 & -u_5 e^{-\alpha} & \partial\gamma & 0 \end{pmatrix},\quad B=\begin{pmatrix} 0 & 0 & 1 & 0 & 0 \cr e^\alpha & 0 & 0 & 0 & 0 \cr 0 & 0 & \bar{\partial}\alpha & v_4 & v_5 \cr 0 & -v_4 e^{-\alpha} & 0 & 0 & \bar{\partial}\bar{\gamma} \cr 0 & -v_5 e^{-\alpha} & 0 & -\bar{\partial}\bar{\gamma} & 0 \end{pmatrix},
\end{equation}
where $u_i$ and $v_i$ are defined to be
\begin{eqnarray}
u_4=+p(z) \cos \bar{\gamma}(z,\bar{z}), &\quad& v_4=-\bar{p}(\bar{z}) \cos \gamma(z,\bar{z}), \\
u_5=-p(z) \sin \bar{\gamma}(z,\bar{z}), &\quad& v_5=-\bar{p}(\bar{z}) \sin \gamma(z,\bar{z}).
\end{eqnarray}
Defining a new field $\beta(z,\bar{z}) \equiv \gamma(z,\bar{z})+\bar{\gamma}(z,\bar{z})$, the zero curvature condition (\ref{eqn427}) implies the equations of motion for the scalar fields
\begin{eqnarray}
&&\partial \bar{\partial} \alpha-e^\alpha+p(z)\bar{p}(\bar{z})e^{-\alpha} \cos \beta=0, \\
&&\partial \bar{\partial} \beta-p(z)\bar{p}(\bar{z})e^{-\alpha} \sin \beta=0.
\end{eqnarray}
After the shift (\ref{change1}) and the change of variables (\ref{change2}), the equations of motion become
\begin{eqnarray}
&&\partial_w \bar{\partial}_{\bar{w}} \hat{\alpha}-e^{\hat{\alpha}}+e^{-\hat{\alpha}}\cos\beta=0, \label{toda1} \\
&&\partial_w \bar{\partial}_{\bar{w}} \beta-e^{-\hat{\alpha}}\sin\beta=0. \label{toda2}
\end{eqnarray}
This is called the $B_2$ Toda system.

\section{Series solution to the sinh-Gordon equation}

We begin our discussion with the application of the method to the sinh-Gordon equation (\ref{sinh}). In this case there exists a powerful multi-integral representation for the radially symmetric solution which emerged from studies of monopole solutions in self-dual Yang-Mills theory and also matrix model integrals. We describe this exact representation in the appendix A. Since we do not posses analogous (integral) representations in the more general Toda case (even though they might definitely exist), we will consider an approximate, series solution to the equations. For a rotationally invariant solution of the sinh-Gordon equation we have the radial Laplacian
\begin{equation}
\partial_w \bar{\partial}_{\bar{w}}={1 \over \rho}{d \over d\rho}\rho{d \over d\rho}={d^2 \over d\rho^2}+{1 \over \rho}{d \over d\rho},
\end{equation}
where $\rho=2\vert w \vert$. One can either expand the equation in series near the origin ($\rho=0$) or at infinity ($\rho=\infty$).

\subsection{Expansion at $\rho=0$}

The sinh-Gordon solution that one seeks, apart from being radially symmetric, also has a well specified singularity at the origin. This singularity is associated with the fact that as we have seen, the AdS$_3$ system consists of a coupled set of equations where after a shift of the field one finds the decoupled sinh-Gordon equation. The shift is given by the conformal (and anti-conformal) fields $p(z)$ and $\bar{p}(\bar{z})$, the degree of the associated conformal mapping then specifies the degree of the singularity which happens as follows.

Consider the regular polygon case \cite{Alday:2009ga} where
\begin{equation}
p(z)=z^{n-2}, \qquad \bar{p}(\bar{z})=\bar{z}^{n-2},
\label{poly}
\end{equation}
the shift (\ref{change1}) becomes
\begin{equation}
\hat{\alpha}=\alpha-{n-2 \over 2}\ln z\bar{z} \sim \alpha-{n-2 \over n}\ln w\bar{w}.
\end{equation}
Furthermore, one expects $\alpha$ to be regular everywhere. Therefore, $\hat{\alpha}$ should have the singularity near $\rho=0$ of the form
\begin{equation}
\hat{\alpha}=-2{n-2 \over n}\ln \rho+\cdots
\end{equation}

Now we are ready to expand the solution near the origin in series for small $\rho$. Denoting $\zeta \equiv (n-2)/n$, we can write the series solution near $\rho=0$ as
\begin{equation}
\hat{\alpha}=-2\zeta\ln\rho+\ln\alpha_0+\sum_{l,k=0 \atop (l,k) \neq (0,0)}^\infty C_{l,k} \rho^{2\gamma_{l,k}},
\label{ansatzsinh}
\end{equation}
where $\gamma_{l,k} \equiv (l+k)+\zeta(l-k)$ and $\alpha_0$ is a constant. Plugging the ansatz (\ref{ansatzsinh}) into the equation of motion (\ref{sinh}), one finds the recursion relation for the coefficients $C_{l,k}$ as
\begin{equation}
4\sideset{}{'}\sum_{l,k} C_{l,k} \gamma_{l,k}^2 \rho^{2\gamma_{l,k}}-\alpha_0 \rho^{2\gamma_{0,1}}\exp[\sideset{}{'}\sum_{l,k} C_{l,k} \rho^{2\gamma_{l,k}}]+\alpha_0^{-1} \rho^{2\gamma_{1,0}}\exp[-\sideset{}{'}\sum_{l,k} C_{l,k} \rho^{2\gamma_{l,k}}]=0,
\end{equation}
where the prime means $(l,k) \neq (0,0)$. The first few coefficients are easily solved to be
\begin{eqnarray}
C_{0,1}&=&{\alpha_0 \over 4(1-\zeta)^2}, \\
C_{1,0}&=&-{1 \over 4\alpha_0(1+\zeta)^2}, \\
C_{1,1}&=&{\zeta \over 16(1-\zeta^2)^2}.
\end{eqnarray}

\subsection{Expansion at $\rho=\infty$}

The essence of the method that we will be following \cite{Manton:1978gf} is to perform an expansion at large distance and then impose the short distance boundary condition (in this case the singularity condition). Even though this represents an extrapolation of the series solution from large all the way to small distance, the technique was known to give excellent results in the case of nonlinear soliton solutions (see appendix B).

For the present problem we impose the natural boundary condition at infinity
\begin{equation}
\hat{\alpha} \to 0 \qquad {\rm at}~~\rho=\infty,
\label{bc2}
\end{equation}
and develop the series expansion of the field $\hat{\alpha}$ at $\rho=\infty$ with the form
\begin{equation}
\hat{\alpha}=\sum_{odd~n} \alpha_n.
\end{equation}
The first order equation reads
\begin{equation}
{1 \over \rho}{d \over d\rho}\rho{d \over d\rho}\alpha_1-2\alpha_1=0.
\end{equation}
The solution satisfying the boundary condition (\ref{bc2}) is
\begin{equation}
\alpha_1=a K_0(\tilde{\rho}),
\end{equation}
where $\tilde{\rho} \equiv \sqrt{2}\rho$, $a$ is an integration constant and $K_0(\tilde{\rho})$ is the modified Bessel function of the second kind. The third order equation reads
\begin{equation}
{1 \over \rho}{d \over d\rho}\rho{d \over d\rho}\alpha_3-2\alpha_3-{1 \over 3}\alpha_1^3=0.
\end{equation}
Using the Green's function, we find the solution as a double integral
\begin{equation}
\alpha_3=K_0(\tilde{\rho})\int_{\tilde{\rho}}^\infty {d\tilde{\rho}' \over \tilde{\rho}' K_0^2(\tilde{\rho}')}\int_{\tilde{\rho}'}^\infty {1 \over 6} a^3 K_0^4(\tilde{\rho}'')\tilde{\rho}'' d\tilde{\rho}'' .
\end{equation}
The fifth order equation reads
\begin{equation}
{1 \over \rho}{d \over d\rho}\rho{d \over d\rho}\alpha_5-2\alpha_5-\alpha_1^2 \alpha_3-{1 \over 60}\alpha_1^5=0.
\end{equation}
The solution can be written as
\begin{equation}
\alpha_5=K_0(\tilde{\rho})\int_{\tilde{\rho}}^\infty {d\tilde{\rho}' \over \tilde{\rho}' K_0^2(\tilde{\rho}')}\int_{\tilde{\rho}'}^\infty S_5^\alpha(\tilde{\rho}'') K_0(\tilde{\rho}'')\tilde{\rho}'' d\tilde{\rho}'' ,
\end{equation}
where the source term is
\begin{equation}
S_5^\alpha(\tilde{\rho}'')={1 \over 2}\alpha_1^2 \alpha_3+{1 \over 120}\alpha_1^5 \Bigr|_{\tilde{\rho}''}.
\end{equation}

Now we perform the important step of confronting the series solution generated above with the short distance boundary requirement 
given by the singularity at $\rho=0$. We find the following relation
\begin{equation}
{\hat{\alpha} \over -\ln\rho} \sim a+{\pi^2 \over 96}a^3+{3\pi^4 a^5 \over 10240}+\cdots=2{n-2 \over n},
\label{series}
\end{equation}
where we have used the asymptotic expansion of the modified Bessel function for small $\rho$: $K_0(\rho) \sim -\ln\rho$. This equation is sufficient to determine the free constant appearing from the leading (first) term of the series. Even though the series (\ref{series}) can be summed up exactly \cite{Zamolodchikov:1994uw}, we will demonstrate that the first few terms will already give a good approximation to the area.

The string worldsheet area can be written as
\begin{eqnarray}
A&=&\int d\tau d\sigma \Bigl({1 \over 2}\partial_\mu Y \partial^\mu Y\Bigr)=\int d^2 z e^\alpha \cr
&=&\int d^2 w e^{\hat{\alpha}}=\int_\Sigma d^2 w + \int d^2 w (e^{\hat{\alpha}}-1).
\end{eqnarray}
The first piece is infinite and can be regularized using the boundary polygon $\Sigma$, while the second piece is finite, which can be denoted as
\begin{equation}
A_{sinh}=\int d^2 w (e^{\hat{\alpha}}-1)={\pi n \over 2} \int_0^\infty \rho d\rho (e^{\hat{\alpha}}-1),
\end{equation}
where we considered a factor of $n/2$ since one round in the $z$ plane will make $n/2$ rounds in the $w$ plane. Using the expansion, the finite piece of area can be further written as
\begin{eqnarray}
A_{sinh}&=&{\pi n \over 4} \int_0^\infty \tilde{\rho} d \tilde{\rho} (e^{\hat{\alpha}}-1) \\
&=&{\pi n \over 4}\int_0^\infty \tilde{\rho} d\tilde{\rho} (\alpha_1+{1 \over 2}\alpha_1^2+\alpha_3+{1 \over 6}\alpha_1^3+\alpha_1 \alpha_3+{1 \over 24}\alpha_1^4+ \cr
& & \qquad \qquad \qquad +\alpha_5+{1 \over 2}\alpha_1^2 \alpha_3+{1 \over 120}\alpha_1^5+\cdots).
\end{eqnarray}
If we include up to the third order terms, the coefficient $a$ is determined by the equation
\begin{equation}
a+{\pi^2 \over 96}a^3=2{n-2 \over n}.
\end{equation}
The area up to third order terms is
\begin{equation}
A_{sinh}^{(3)}={\pi n \over 4}(a+{1 \over 4}a^2+0.102808 a^3+\cdots).
\end{equation}
We compare our result to the known exact answer \cite{Alday:2009ga} (see appendix A for its derivation)
\begin{equation}
A_{exact}={\pi \over 4n}(3n^2-8n+4).
\label{exact}
\end{equation}
The comparison and numerical results are summarized in Table \ref{table1}.

If we include terms up to fifth order in our expansion, the coefficient $a$ is determined by the equation
\begin{equation}
a+{\pi^2 \over 96}a^3+{3\pi^4 a^5 \over 10240}=2{n-2 \over n}.
\end{equation}
The area up to fifth order terms is
\begin{equation}
A_{sinh}^{(5)}={\pi n \over 4}(a+{1 \over 4}a^2+0.102808 a^3+0.0514042 a^4+0.0285378 a^5+\cdots).
\end{equation}
Comparing to the exact answer (\ref{exact}), we list the numerical results in Table \ref{table1}.

\begin{table}[h]
\begin{center}
\begin{tabular}{|c|c|c|c|c|c|}
\hline
 & $A_{exact}$ & $A_{sinh}^{(3)}$ & Error$^{(3)}$ & $A_{sinh}^{(5)}$ & Error$^{(5)}$  \\
\hline
$2n=6$ & 1.83260 & 1.81188  & 1.13\% & 1.82983 & 0.15\%  \\
\hline
$2n=8$ & 3.92699 & 3.80629 & 3.07\% & 3.89526  & 0.81\% \\
\hline
$2n=10$ & 6.12611 & 5.84269 & 4.63\% & 6.02938  & 1.58\% \\
\hline
$2n=12$ & 8.37758 & 7.89331 & 5.78\% & 8.18848  & 2.26\% \\
\hline
\end{tabular}
\caption{Comparison of the estimation with the exact results.}
\label{table1}
\end{center}
\end{table}

Clearly for small $n$, just a few orders in the expansion that we generate are capable of producing a solution giving the area close to the exact one. One also notices that the error increases when $n$ gets larger.

\section{Series solution of the AdS$_4$ system}

We now proceed to demonstrate the applicability of this solution generating technique to the next case of $B_2$ Toda system, which appears in the Pohlmeyer reduction of AdS$_4$ string equations. The procedure parallels the calculation of previous section.

\subsection{Expansion at $\rho=0$}

We now have one more scalar field and the coupled system of nonlinear equations of the Toda system. Regarding behavior at the origin we again expect $\alpha,\beta$ to be regular everywhere. If we choose the same polynomial form (\ref{poly}) for the conformal map, the series expansion near $\rho=0$ for the Toda fields reads
\begin{eqnarray}
\hat{\alpha}&=&-2\zeta\ln\rho+\ln\alpha_0+\sideset{}{'}\sum_{l,k=0}^\infty C_{l,k}^\alpha \rho^{2\gamma_{l,k}}, \label{anstoda1} \\
\beta&=&\beta_0+\sideset{}{'}\sum_{l,k=0}^\infty C_{l,k}^\beta \rho^{2\gamma_{l,k}}, \label{anstoda2}
\end{eqnarray}
where $\alpha_0,\beta_0$ are constants. Plugging the ansatz (\ref{anstoda1}, \ref{anstoda2}) into the equations of motion (\ref{toda1}, \ref{toda2}), we get the recursion relations for the coefficients $C_{l,k}^\alpha,C_{l,k}^\beta$ as
\begin{multline}
4\sideset{}{'}\sum_{l,k} C_{l,k}^\alpha \gamma_{l,k}^2 \rho^{2\gamma_{l,k}}-\alpha_0 \rho^{2\gamma_{0,1}}\exp[\sideset{}{'}\sum_{l,k} C_{l,k}^\alpha \rho^{2\gamma_{l,k}}]+\alpha_0^{-1} \rho^{2\gamma_{1,0}} \exp[-\sideset{}{'}\sum_{l,k} C_{l,k}^\alpha \rho^{2\gamma_{l,k}}] \cr
\times \Bigl(\cos\beta_0\cos[\sideset{}{'}\sum_{l,k} C_{l,k}^\beta \rho^{2\gamma_{l,k}}]-\sin\beta_0\sin[\sideset{}{'}\sum_{l,k} C_{l,k}^\beta \rho^{2\gamma_{l,k}}]\Bigr)=0,
\end{multline}
\begin{multline}
4\sideset{}{'}\sum_{l,k} C_{l,k}^\beta \gamma_{l,k}^2 \rho^{2\gamma_{l,k}}-\alpha_0^{-1} \rho^{2\gamma_{1,0}} \exp[-\sideset{}{'}\sum_{l,k} C_{l,k}^\alpha \rho^{2\gamma_{l,k}}] \cr
\times \Bigl(\sin\beta_0\cos[\sideset{}{'}\sum_{l,k} C_{l,k}^\beta \rho^{2\gamma_{l,k}}]+\cos\beta_0\sin[\sideset{}{'}\sum_{l,k} C_{l,k}^\beta \rho^{2\gamma_{l,k}}]\Bigr)=0.
\end{multline}
The first few coefficients are solved to be
\begin{eqnarray}
C_{0,1}^\alpha &=& {\alpha_0 \over 4(1-\zeta)^2}, \\
C_{0,1}^\beta &=& 0, \\
C_{1,0}^\alpha &=& -{\cos \beta_0 \over \alpha_0}{1 \over 4(1+\zeta)^2}, \\
C_{1,0}^\beta &=& {\sin \beta_0 \over \alpha_0}{1 \over 4(1+\zeta)^2}, \\
C_{1,1}^\alpha &=& \cos\beta_0 {\zeta \over 16(1-\zeta^2)^2},
\end{eqnarray}
\begin{eqnarray}
C_{1,1}^\beta &=& -\sin\beta_0 {1 \over 64(1-\zeta)^2}.
\end{eqnarray}

\subsection{Expansion at $\rho=\infty$}

We impose the boundary condition for the Toda system
\begin{equation}
\hat{\alpha},\beta \to 0 \qquad {\rm at}~~\rho=\infty,
\label{bc}
\end{equation}
and generate the long distance expansion in the form
\begin{equation}
\hat{\alpha}=\sum_{n=1}^\infty \alpha_n, \qquad \beta=\sum_{n=1}^\infty \beta_n.
\end{equation}
The first order equations are decoupled
\begin{eqnarray}
&&{1 \over \rho}{d \over d\rho}\rho{d \over d\rho}\alpha_1-2\alpha_1=0, \\
&&{1 \over \rho}{d \over d\rho}\rho{d \over d\rho}\beta_1-\beta_1=0,
\end{eqnarray}
with the solution satisfying the boundary condition (\ref{bc}) as
\begin{eqnarray}
&&\alpha_1(\tilde{\rho})=a K_0(\tilde{\rho}), \\
&&\beta_1(\rho)=b K_0(\rho),
\end{eqnarray}
where $\tilde{\rho} \equiv \sqrt{2}\rho$ and $a,b$ are the integration constants. The second order equations are
\begin{eqnarray}
&&{1 \over \rho}{d \over d\rho}\rho{d \over d\rho}\alpha_2-2\alpha_2-{1 \over 2}\beta_1^2=0, \\
&&{1 \over \rho}{d \over d\rho}\rho{d \over d\rho}\beta_2-\beta_2+\alpha_1 \beta_1=0.
\end{eqnarray}
Using the Green's function, we can treat the third terms as sources and the solutions can be written in terms of double integrals
\begin{eqnarray}
&&\alpha_2(\tilde{\rho})=+K_0(\tilde{\rho}) \int_{\tilde{\rho}}^\infty {d\tilde{\rho}' \over \tilde{\rho}' K_0^2(\tilde{\rho}')} \int_{\tilde{\rho}'}^\infty {1 \over 4}b^2 K_0^2(\tilde{\rho}'' / \sqrt{2}) K_0(\tilde{\rho}'') \tilde{\rho}'' d\tilde{\rho}'' , \\
&&\beta_2(\rho)=-K_0(\rho)\int_\rho^\infty {d\rho' \over \rho' K_0^2(\rho')} \int_{\rho'}^\infty a b K_0(\sqrt{2} \rho'')K_0^2(\rho'') \rho'' d\rho'' .
\end{eqnarray}
The third order equations are
\begin{eqnarray}
&&{1 \over \rho}{d \over d\rho}\rho{d \over d\rho}\alpha_3-2\alpha_3-{1 \over 3}\alpha_1^3+{1 \over 2}\alpha_1 \beta_1^2-\beta_1 \beta_2=0, \\
&&{1 \over \rho}{d \over d\rho}\rho{d \over d\rho}\beta_3-\beta_3+{1 \over 6}\beta_1^3-{1 \over 2}\alpha_1^2 \beta_1+\alpha_1\beta_2+\alpha_2\beta_1=0.
\end{eqnarray}
The solutions are
\begin{eqnarray}
&&\alpha_3(\tilde{\rho})=+K_0(\tilde{\rho}) \int_{\tilde{\rho}}^\infty {d\tilde{\rho}' \over \tilde{\rho}' K_0^2(\tilde{\rho}')} \int_{\tilde{\rho}'}^\infty S_3^\alpha(\tilde{\rho}'') K_0(\tilde{\rho}'') \tilde{\rho}'' d\tilde{\rho}'' , \\
&&\beta_3(\rho)=-K_0(\rho)\int_\rho^\infty {d\rho' \over \rho' K_0^2(\rho')} \int_{\rho'}^\infty S_3^\beta(\rho'') K_0(\rho'') \rho'' d\rho'' ,
\end{eqnarray}
where the source terms are
\begin{eqnarray}
&&S_3^\alpha(\tilde{\rho}'')={1 \over 6}\alpha_1^3-{1 \over 4}\alpha_1\beta_1^2+{1 \over 2}\beta_1 \beta_2 \Bigr|_{\tilde{\rho}''}, \\
&&S_3^\beta(\rho'')={1 \over 6}\beta_1^3-{1 \over 2}\alpha_1^2 \beta_1+\alpha_1\beta_2+\alpha_2\beta_1 \Bigr|_{\rho''}.
\end{eqnarray}

Now we pull back the solution to $\rho=0$, up to third order, we get
\begin{eqnarray}
{\hat{\alpha} \over -\ln\rho}&=&a+{\pi \over 8}b^2+{\pi^2 \over 96}a^3-{\pi^2 \over 32}a b^2, \\
{\beta \over -\ln\rho}&=&b-{\pi \over 8}a b-{7\pi^2 \over 384}b^3+{3\pi^2 \over 128}a^2 b.
\end{eqnarray}
The coefficients $a,b$ are determined by the following equations
\begin{eqnarray}
a+{\pi \over 8}b^2+{\pi^2 \over 96}a^3-{\pi^2 \over 32}a b^2&=&2{n-2 \over n}, \\
b-{\pi \over 8}a b-{7\pi^2 \over 384}b^3+{3\pi^2 \over 128}a^2 b&=&0.
\end{eqnarray}
The finite piece of the area up to third order can be written as
\begin{eqnarray}
A_{Toda}&=&{\pi n \over 4}\int_0^\infty \tilde{\rho} d\tilde{\rho} (\alpha_1+\alpha_2+{1 \over 2}\alpha_1^2+\alpha_3+\alpha_1\alpha_2+{1 \over 6}\alpha_1^3+\cdots), \\
&=&{\pi n \over 4}(a+{1 \over 4}a^2+0.142699 b^2+0.102808 a^3+0.0842739 a b^2+\cdots),
\end{eqnarray}
which is minimized at
\begin{equation}
a+{\pi^2 \over 96}a^3=2{n-2 \over n}, \qquad b=0,
\end{equation}
for small $n \le n_c$. When $n > n_c$, $A_{Toda}$ is minimized at nonzero $b$. However, the error of the approximation is also expected to increase. The value of $n_c$ depends on the order of expansion. Therefore, this effect might be artifact of the approximation itself. One might have an argument why $b=0$ is the preferred minimal area solution. The contribution to the area mostly comes near $\rho=0$ with the singularity. Consider the asymptotic behavior
\begin{equation}
\hat{\alpha} \sim -2{n-2 \over n}\ln \rho,
\end{equation}
we have
\begin{equation}
e^{-\hat{\alpha}} \sim \rho^{2(n-2)/n}.
\end{equation}
For $n \ge 3$, we find $e^{-\hat{\alpha}} \to 0$ as $\rho \to 0$. Furthermore, $\cos\beta$ and $\sin\beta$ are finite, they will not contribute much because of the vanishing factor $e^{-\hat{\alpha}}$.

Concluding this section, we have generated a one parameter approximate solution to the AdS$_4$ string equations. For finding the minimal area one can perform a minimization with respect to the free parameter and obtain an approximate result with higher and higher accuracy.

{\bf Note added:} In a very recent publication \cite{Alday:2009dv}, Alday, Gaiotto and Maldacena considered minimal surfaces in AdS$_5$ with null polygonal boundary conditions and argued it is relevant to continue from the Lorentzian AdS space with (3,1) signature to (2,2) signature. In that case of AdS$_4$, the scalar fields $\hat{\alpha}$ and $\beta$ can both have logarithmic singularities near $\rho=0$ as
\begin{equation}
\hat{\alpha}=-2\zeta\ln\rho+c_\alpha+\cdots, \qquad \beta=-4\zeta\ln\rho+c_\beta+\cdots,
\end{equation}
where the constants $c_\alpha,c_\beta$ can be determined by the boundary conditions at infinity (\ref{bc}). This analytic continuation can be achieved by taking $\beta \to i\beta$ of (\ref{toda1}-\ref{toda2}). For the large $\rho$ solution, all one needs to do is to take $b \to i b$. Therefore, the matching conditions for the singularities give
\begin{eqnarray}
a-{\pi \over 8}b^2+{\pi^2 \over 96}a^3+{\pi^2 \over 32}a b^2&=&2{n-2 \over n}, \\
b-{\pi \over 8}a b+{7\pi^2 \over 384}b^3+{3\pi^2 \over 128}a^2 b&=&4{n-2 \over n},
\end{eqnarray}
and the area is given by
\begin{eqnarray}
A_{Toda}^{++}&=&{\pi n \over 4}\int_0^\infty \tilde{\rho} d\tilde{\rho} (\alpha_1+\alpha_2+{1 \over 2}\alpha_1^2+\alpha_3+\alpha_1\alpha_2+{1 \over 6}\alpha_1^3+\cdots), \\
&=&{\pi n \over 4}(a+{1 \over 4}a^2-0.142699 b^2+0.102808 a^3-0.0842739 a b^2+\cdots).
\end{eqnarray}
In the case of pentagon where $2n=5$, we find $A_{\rm pentagon}^{++} \approx 0.97849$. Comparing with the shooting method result \cite{Alday:2009dv}, the difference is about 17\%. Better accuracy can be obtained by adding more terms.

\section{Conclusion}

We have described an approximate method for construction of Euclidean instanton type solutions of Toda equations associated with 
minimal surfaces in AdS$_d$. The method is based on a series solution at large distance with a nontrivial matching at short distance which results in specification of constants appearing in the expansion. In the case of sinh-Gordon, the finite piece of area calculated up to a few terms is seen to be very close to the exact result. The procedure is then extended and demonstrated to be applicable for the generalized Toda systems associated with AdS$_d$ strings. We treated in detail the $B_2$ Toda case corresponding to AdS$_4$.

Our calculation concerns only one piece of the solution for the Plateau problem in AdS$_4$, the contribution associated with the singular instanton of the Toda system $A_{Toda}$. The contribution given by the radial (instanton) solution does not depend on the cross ratios. The most nontrivial part in the construction of the string solution with null polygonal boundaries is the matching of polygon boundary conditions as done by Alday and Maldacena \cite{Alday:2009ga}.  It remains a challenge to evaluate the complete expression with the terms associated with cross ratios.

\acknowledgments

We would like to thank J. Avan, C. Kalousios and C. Vergu for comments and discussions. This work is supported by the Department of Energy under contract DE-FG02-91ER40688.

\appendix

\section{Exact solution to the sinh-Gordon equation}

In this appendix, we describe the exact integral representation for the radially symmetric solution to the sinh-Gordon equation and derive the exact result (\ref{exact}). One notices the exact solution to the sinh-Gordon equation (\ref{sinh}) following \cite{Zamolodchikov:1994uw} with a very nontrivial integral represention \cite{TTWu}
\begin{equation}
\hat{\alpha}(\rho\vert\lambda)=W(\rho\vert\lambda)-W(\rho\vert-\lambda),
\end{equation}
where $W(\rho\vert\lambda)$ has the expansion
\begin{equation}
W(\rho\vert\lambda)=2\sum_{k=1}^\infty {\lambda^k \over k}W_k(\rho)=2\sum_{k=1}^\infty {\lambda^k \over k}\int_0^\infty \prod_{i=1}^k {2e^{-\sqrt{2}\rho\cosh[\ln x_i]} \over x_i+x_{i+1}} dx_i,
\label{seriesexact}
\end{equation}
with $x_{k+1}=x_1$. One can verify order by order of the $\lambda$-expansion that the following relation is correct
\begin{equation}
{1 \over \rho}{d \over d\rho}\rho{d \over d\rho} W(\rho\vert\lambda)=e^{\hat{\alpha}}-1.
\end{equation}
Therefore, the finite piece of the area $A_{sinh}$ can be calculated as
\begin{equation}
A_{sinh}={\pi n \over 2}\int_0^\infty \rho d\rho (e^{\hat{\alpha}}-1)={\pi n \over 2} \Bigl( \rho {d W \over d\rho} \Bigr) \Bigr|_{\rho=0}^{\rho=\infty}.
\end{equation}
Thus one only needs to look at the series (\ref{seriesexact}) at large and small $\rho$. The large $\rho$ expansion shows
\begin{equation}
W(\rho\vert\lambda)=4\lambda K_0(\tilde{\rho})+16\lambda^2\Bigl[-\ln \tilde{\rho} \int_{\tilde{\rho}}^\infty K_0^2(\tilde{\rho}')\tilde{\rho}' d\tilde{\rho}' +\int_{\tilde{\rho}}^\infty K_0^2(\tilde{\rho}')\tilde{\rho}' \ln \tilde{\rho}' d\tilde{\rho}' \Bigr]+\cdots
\end{equation}
where $\tilde{\rho} \equiv \sqrt{2}\rho$. Asymptotically, $K_0(\rho)=e^{-\rho}/\sqrt{\rho}$, so that
\begin{equation}
\rho {d W \over d\rho} \sim -\sqrt{\rho} e^{-\sqrt{2}\rho} \to 0, \qquad {\rm as} \quad \rho \to \infty.
\end{equation}
On the other hand, the small $\rho$ expansion of the series (\ref{seriesexact}) gives
\begin{equation}
W(\rho\vert\lambda)=-{\zeta(\zeta+2) \over 2}\ln \rho+\cdots
\label{easy}
\end{equation}
where $\zeta$ is related to $\lambda$ by
\begin{equation}
2\pi\lambda=\sin{\pi\zeta \over 2}.
\end{equation}
From (\ref{easy}), one can easily evaluate
\begin{equation}
\rho {d W \over d\rho} \sim -{\zeta(\zeta+2) \over 2}, \qquad {\rm as} \quad \rho \to 0.
\end{equation}
Recalling $\zeta \equiv (n-2)/n$, we get the exact result
\begin{equation}
A_{sinh}={\pi \over 4 n}(3n^2-8n+4).
\end{equation}

\section{Series expansion of the kink solution}

In this appendix, we discuss the 1+1 dimensional $\varphi^4$ theory as an illustrative example for the method employed in \cite{Manton:1978gf}. The static Lagrangian reads
\begin{equation}
L=\int dx \Bigl( -{1 \over 2}\varphi_{x}^2-{1 \over 4}(1-\varphi^2)^2 \Bigr),
\end{equation}
from which the equation of motion can be derived as
\begin{equation}
\varphi_{xx}+\varphi(1-\varphi^2)=0.
\label{eomphi4}
\end{equation}
Suppose a series solution of the scalar field at $x=-\infty$,
\begin{equation}
\varphi(x)=-1+\sum_{n=1}^\infty c_n e^{n \sqrt{2} x}.
\end{equation}
The equation of motion (\ref{eomphi4}) gives the recursion relation
\begin{equation}
2\sum_n c_n n^2 e^{n \sqrt{2} x}-2\sum_n c_n e^{n \sqrt{2} x}+3\sum_{m,l} c_m c_l e^{(m+l) \sqrt{2} x}-\sum_{m,l,k} c_m c_l c_k e^{(m+l+k) \sqrt{2} x}=0.
\end{equation}
Comparing the coefficients term by term, we find $c_1$ is undetermined and
\begin{equation}
c_2=-{1 \over 2} c_1^2, \quad c_3={1 \over 4} c_1^3, \quad c_4=-{1 \over 8} c_1^4, \quad \cdots
\end{equation}
One can sum up the series and get
\begin{equation}
\varphi=-1+c_1 e^{\sqrt{2}x}-{1 \over 2}c_1^2 e^{2\sqrt{2}x}+{1 \over 4}c_1^3 e^{3\sqrt{2}x}-{1 \over 8}c_1^4 e^{4\sqrt{2}x}+\cdots ={{1 \over 2}c_1 e^{\sqrt{2}x}-1 \over {1 \over 2}c_1 e^{\sqrt{2}x}+1}.
\end{equation}
Impose the boundary condition at origin
\begin{equation}
\varphi(0)=0,
\end{equation}
we get $c_1=2$ so that the solution reads
\begin{equation}
\varphi={e^{\sqrt{2}x}-1 \over e^{\sqrt{2}x}+1}=\tanh {x \over \sqrt{2}},
\end{equation}
which is exactly the kink solution.


\begin{thebibliography}{99}

\bibitem{Gubser:2002tv}
  S.~S.~Gubser, I.~R.~Klebanov and A.~M.~Polyakov,
  Nucl.\ Phys.\  B {\bf 636}, 99 (2002)
  [arXiv:hep-th/0204051].

\bibitem{Beisert:2006ez}
  N.~Beisert, B.~Eden and M.~Staudacher,
  J.\ Stat.\ Mech.\  {\bf 0701}, P021 (2007)
  [arXiv:hep-th/0610251].

\bibitem{Tseytlin:2003ii}
  A.~A.~Tseytlin,
  arXiv:hep-th/0311139.

  A.~A.~Tseytlin,
  arXiv:hep-th/0409296.

  J.~Plefka,
  Living Rev.\ Rel.\  {\bf 8}, 9 (2005)
  [arXiv:hep-th/0507136].

\bibitem{Kruczenski:2004wg}
  M.~Kruczenski,
  JHEP {\bf 0508}, 014 (2005)
  [arXiv:hep-th/0410226].

  M.~Kruczenski and A.~A.~Tseytlin,
  Phys.\ Rev.\  D {\bf 77}, 126005 (2008)
  [arXiv:0802.2039 [hep-th]].

  N.~Dorey,
  Acta Phys.\ Polon.\  B {\bf 39}, 3081 (2008)
  [arXiv:0805.4387 [hep-th]].

  N.~Dorey and M.~Losi,
  arXiv:0812.1704 [hep-th].

\bibitem{Alday:2007hr}
  L.~F.~Alday and J.~M.~Maldacena,
  JHEP {\bf 0706}, 064 (2007)
  [arXiv:0705.0303 [hep-th]].

  L.~F.~Alday and J.~Maldacena,
  JHEP {\bf 0711}, 068 (2007)
  [arXiv:0710.1060 [hep-th]].

\bibitem{Alday:2008yw}
  L.~F.~Alday and R.~Roiban,
  Phys.\ Rept.\  {\bf 468}, 153 (2008)
  [arXiv:0807.1889 [hep-th]].

  J.~M.~Drummond, G.~P.~Korchemsky and E.~Sokatchev,
  Nucl.\ Phys.\  B {\bf 795}, 385 (2008)
  [arXiv:0707.0243 [hep-th]].

  J.~M.~Drummond, J.~Henn, G.~P.~Korchemsky and E.~Sokatchev,
  Nucl.\ Phys.\  B {\bf 795}, 52 (2008)
  [arXiv:0709.2368 [hep-th]].

  J.~M.~Drummond, J.~Henn, G.~P.~Korchemsky and E.~Sokatchev,
  Nucl.\ Phys.\  B {\bf 815}, 142 (2009)
  [arXiv:0803.1466 [hep-th]].

  A.~Brandhuber, P.~Heslop and G.~Travaglini,
  Nucl.\ Phys.\  B {\bf 794}, 231 (2008)
  [arXiv:0707.1153 [hep-th]].

  C.~Anastasiou, A.~Brandhuber, P.~Heslop, V.~V.~Khoze, B.~Spence and G.~Travaglini,
  JHEP {\bf 0905}, 115 (2009)
  [arXiv:0902.2245 [hep-th]].

  A.~Brandhuber, P.~Heslop, V.~V.~Khoze and G.~Travaglini,
  arXiv:0910.4898 [hep-th].

\bibitem{Bern:2005iz}
  Z.~Bern, L.~J.~Dixon and V.~A.~Smirnov,
  Phys.\ Rev.\  D {\bf 72}, 085001 (2005)
  [arXiv:hep-th/0505205].

\bibitem{Anastasiou:2003kj}
  C.~Anastasiou, Z.~Bern, L.~J.~Dixon and D.~A.~Kosower,
  Phys.\ Rev.\ Lett.\  {\bf 91}, 251602 (2003)
  [arXiv:hep-th/0309040].

  Z.~Bern, J.~J.~M.~Carrasco, H.~Johansson and D.~A.~Kosower,
  Phys.\ Rev.\  D {\bf 76}, 125020 (2007)
  [arXiv:0705.1864 [hep-th]].

  Z.~Bern, L.~J.~Dixon, D.~A.~Kosower, R.~Roiban, M.~Spradlin, C.~Vergu and A.~Volovich,
  Phys.\ Rev.\  D {\bf 78}, 045007 (2008)
  [arXiv:0803.1465 [hep-th]].

  F.~Cachazo, M.~Spradlin and A.~Volovich,
  Phys.\ Rev.\  D {\bf 78}, 105022 (2008)
  [arXiv:0805.4832 [hep-th]].

  M.~Spradlin, A.~Volovich and C.~Wen,
  Phys.\ Rev.\  D {\bf 78}, 085025 (2008)
  [arXiv:0808.1054 [hep-th]].

  C.~Vergu,
  arXiv:0908.2394 [hep-th].

\bibitem{Mironov:2007qq}
  A.~Mironov, A.~Morozov and T.~N.~Tomaras,
  JHEP {\bf 0711}, 021 (2007)
  [arXiv:0708.1625 [hep-th]].

  H.~Itoyama, A.~Mironov and A.~Morozov,
  Nucl.\ Phys.\  B {\bf 808}, 365 (2009)
  [arXiv:0712.0159 [hep-th]].

  H.~Itoyama and A.~Morozov,
  Prog.\ Theor.\ Phys.\  {\bf 120}, 231 (2008)
  [arXiv:0712.2316 [hep-th]].

\bibitem{Dobashi:2008ia}
  S.~Dobashi, K.~Ito and K.~Iwasaki,
  JHEP {\bf 0807}, 088 (2008)
  [arXiv:0805.3594 [hep-th]].

  S.~Dobashi and K.~Ito,
  Nucl.\ Phys.\  B {\bf 819}, 18 (2009)
  [arXiv:0901.3046 [hep-th]].

\bibitem{Pohlmeyer}
 K. Pohlmeyer,
 Commun. Math. Phys. {\bf 46}, 207 (1976).

\bibitem{Jevicki:2007aa}
  A.~Jevicki, K.~Jin, C.~Kalousios and A.~Volovich,
  JHEP {\bf 0803}, 032 (2008)
  [arXiv:0712.1193 [hep-th]].

\bibitem{de Vega}
 H.~J.~de~Vega and N.~Sanchez,
 Phys. Rev. D {\bf 47}, 3394 (1993).

\bibitem{Jevicki:2008mm}
  A.~Jevicki and K.~Jin,
  Int.\ J.\ Mod.\ Phys.\  A {\bf 23}, 2289 (2008)
  [arXiv:0804.0412 [hep-th]].

  A.~Jevicki and K.~Jin,
  JHEP {\bf 0906}, 064 (2009)
  [arXiv:0903.3389 [hep-th]].

\bibitem{Berkovits:2008ic}
  N.~Berkovits and J.~Maldacena,
  JHEP {\bf 0809}, 062 (2008)
  [arXiv:0807.3196 [hep-th]].

  L.~F.~Alday, J.~M.~Henn, J.~Plefka and T.~Schuster,
  arXiv:0908.0684 [hep-th].

\bibitem{Dorn:2009kq}
  H.~Dorn, G.~Jorjadze and S.~Wuttke,
  arXiv:0903.0977 [hep-th].

  K.~Sakai and Y.~Satoh,
  JHEP {\bf 0910}, 001 (2009)
  [arXiv:0907.5259 [hep-th]].

  H.~Dorn,
  arXiv:0910.0934 [hep-th].

  S.~Ryang,
  arXiv:0910.4796 [hep-th].

\bibitem{Alday:2009ga}
  L.~F.~Alday and J.~Maldacena,
  arXiv:0903.4707 [hep-th].

  L.~F.~Alday and J.~Maldacena,
  arXiv:0904.0663 [hep-th].

\bibitem{Gaiotto:2008cd}
  D.~Gaiotto, G.~W.~Moore and A.~Neitzke,
  arXiv:0807.4723 [hep-th].

  D.~Gaiotto, G.~W.~Moore and A.~Neitzke,
  arXiv:0907.3987 [hep-th].

\bibitem{TTWu}
 B. M. McCoy, C. A. Tracy and T. T. Wu, J. Math. Phys. {\bf 18}, 1058 (1977).

\bibitem{Zamolodchikov:1994uw}
  A.~B.~Zamolodchikov,
  Nucl.\ Phys.\  B {\bf 432}, 427 (1994)
  [arXiv:hep-th/9409108].

\bibitem{CFYG}
 E. F. Corrigan, D. B. Fairlie, R. G. Yates and P. Goddard, Commun. Math. Phys. {\bf 58}, 223 (1978).

 P. Rossi, Phys. Lett. B {\bf 117}, 72 (1982).

\bibitem{Manton:1978gf}
  N.~S.~Manton,
  Nucl.\ Phys.\  B {\bf 150} (1979) 397.

\bibitem{Barbashov:1982qz}
  B.~M.~Barbashov, V.~V.~Nesterenko and A.~M.~Chervyakov,
  Commun.\ Math.\ Phys.\  {\bf 84}, 471 (1982).

  B.~M.~Barbashov, V.~V.~Nesterenko and A.~M.~Chervyakov,
  Theo.\ Math.\ Phys.\  {\bf 59}, 458 (1984).

  I.~Bakas, Q.~H.~Park and H.~J.~Shin,
  Phys.\ Lett.\  B {\bf 372}, 45 (1996)
  [arXiv:hep-th/9512030].

  J.~L.~Miramontes,
  JHEP {\bf 0810}, 087 (2008)
  [arXiv:0808.3365 [hep-th]].

  M.~Grigoriev and A.~A.~Tseytlin,
  Nucl.\ Phys.\  B {\bf 800}, 450 (2008)
  [arXiv:0711.0155 [hep-th]].

  M.~Grigoriev and A.~A.~Tseytlin,
  Int.\ J.\ Mod.\ Phys.\  A {\bf 23}, 2107 (2008)
  [arXiv:0806.2623 [hep-th]].

  A.~Mikhailov and S.~Schafer-Nameki,
  JHEP {\bf 0805}, 075 (2008)
  [arXiv:0711.0195 [hep-th]].

\bibitem{Alday:2009dv}
  L.~F.~Alday, D.~Gaiotto and J.~Maldacena,
  arXiv:0911.4708 [hep-th].

\end{thebibliography}
\end{document}